\documentclass[a4paper,fleqn,usenatbib,useAMS]{mnras}
\usepackage{graphicx}	% Including figure files
\usepackage{amsmath}	% Advanced maths commands
\usepackage{amssymb}	% Extra maths symbols
\usepackage{multicol}   % Multi-column entries in tables
\usepackage{bm}		    % Bold maths symbols, including upright Greek
\usepackage{pdflscape}	% Landscape pages
\usepackage{array}
\usepackage{tabularx}
\usepackage{rotating}

\usepackage[T1]{fontenc}
\usepackage{ae,aecompl}

\title[Selection of Quasar Candidates]{Efficient Selection of Quasar Candidates Based on Optical and Infrared Photometric Data Using Machine Learning}

\author[X. Jin et~al.]
  {Xin Jin$^{1,2}$, Yanxia Zhang$^{1}$\thanks{Email: zyx@bao.ac.cn}
  , Jingyi Zhang$^{1,2}$, Yongheng Zhao$^1$, \and Xue-bing Wu$^{3,4}$, Dongwei Fan$^1$\\
  $^1$Key Laboratory of Optical Astronomy, National Astronomical
Observatories, Chinese Academy of Sciences, Beijing, 100101, China\\
  $^2$University of Chinese Academy of Sciences, Beijing 100049, China\\
  $^3$Kavli Institute for Astronomy and Astrophysics, Peking University, Beijing 100871, China\\
  $^4$Department of Astronomy, School of Physics, Peking University, Beijing 100871, China}

\date{Accepted XXX. Received YYY; in original form ZZZ}

\pubyear{2018}

\begin{document}
\label{firstpage}
\pagerange{\pageref{firstpage}--\pageref{lastpage}}
\maketitle

% Abstract of the paper
\begin{abstract}
We aim to select quasar candidates based on the two large survey databases, Pan-STARRS and AllWISE. Exploring the distribution of quasars and stars in the color spaces, we find that the combination of infrared and optical photometry is more conducive to select quasar candidates. Two new color criterions (\texttt{yW1W2} and \texttt{iW1zW2}) are constructed to distinguish quasars from stars efficiently. With \texttt{izW1W2}, 98.30\% of star contamination is eliminated, while 99.50\% of quasars are retained, at least to the magnitude limit of our training set of stars. Based on the optical and infrared color features, we put forward an efficient schema to select quasar candidates and high redshift quasar candidates, in which two machine learning algorithms (XGBoost and SVM) are implemented. The XGBoost and SVM classifiers have proven to be very effective with accuracy of $99.46\%$ when \texttt{8Color} as input pattern and default model parameters. Applying the two optimal classifiers to the unknown Pan-STARRS and AllWISE cross-matched data set, a total of 2,006,632 intersected sources are predicted to be quasar candidates given quasar probability larger than 0.5 (i.e. $P_{\mathrm {QSO}}>0.5$). Among them, 1,201,211 have high probability ($P_{\mathrm {QSO}}>0.95$). For these newly predicted quasar candidates, a regressor is constructed to estimate their redshifts. Finally 7,402 $z>3.5$ quasars are obtained. Given the magnitude limitation and site of the LAMOST telescope, part of these candidates will be used as the input catalogue of the LAMOST telescope for follow-up observation, and the rest may be observed by other telescopes.
\end{abstract}

% Select between one and six entries from the list of approved keywords.
% Don't make up new ones.

%%%%%%%%%%%%%%%%%%%%%%%%%%%%%%%%%%%%%%%%%%%%%%%%%%

%%%%%%%%%%%%%%%%% BODY OF PAPER %%%%%%%%%%%%%%%%%%
\section{Introduction}

Quasars are high-luminosity objects that are observed at extremely distant distances, up to 12.9 billion light-years \citep{2011Natur.474..616M}, and they are believed to be powered by the
accretion onto supermassive black holes (SMBHs) in the centers of galaxies \citep{1993ARA&A..31..473A}. By tracing the properties of quasars, we may understand supermassive black holes in massive galaxies, and the coevolution of black holes and their host galaxies \citep{2013ARA&A..51..511K}.
So far, quasars have been found to distribute on a very wide redshift range from 0 to over 7 \citep{2011Natur.474..616M}. High redshift quasars can be taken as important probes for the formation and evolution of structure in the early universe \citep{2006NewAR..50..665F}.

Since the first quasar was discovered in 1963 \citep{1963Natur.197.1040S}, the number of quasars has grown significantly, especially after the implementation of
many wide-field spectroscopical surveys, such as the Two-Degree Field Quasar Redshift Survey (2dF; \citealt{2004MNRAS.349.1397C}), the Sloan Digital Sky Survey (SDSS; \citealt{York_2000}) and Large Sky Area Multi-Object
Fiber Spectroscopic Telescope (LAMOST; \citealt{2012RAA....12.1197C}).

SDSS has conducted several quasar surveys in its different phases.
The quasar program of SDSS-I/II led to the discovery of redshift z > 5 quasars \citep{1999AJ....118....1F} and provided the DR7 quasar catalog (DR7Q; \citealt{2010AJ....139.2360S}), which contains more than 105,000 quasars.
The SDSS-III Baryon Oscillation Spectroscopic Survey (BOSS; \citealt{2012ApJS..199....3R, 2013AJ....145...10D}) has discovered about 270,000 quasars, mostly in the redshift range $2.15-3.5$, which were released in DR12 quasar catalog (DR12Q; \citealt{2017A&A...597A..79P}).
SDSS-IV extended Baryon Oscillation Spectroscopic Survey (eBOSS; \citealt{2015ApJS..221...27M, 2016AJ....151...44D}) concentrates its efforts on the observation of galaxies and in particular quasars. The SDSS-DR14 quasar catalog (DR14Q, \citealt{2018A&A...613A..51P}), which is the first to be released that contains new identifications from eBOSS, contains 526,356 quasars among which 144,046 are new discoveries.

So many quasars have been discovered mainly depending on large photometric sky survey projects (for example, GALEX, SDSS, Pan-STARRS, WISE, 2MASS) and efficient selection of quasar candidates. There are various approaches focusing on targeting quasar candidates, which groups into color-color cut, variability selection and machine learning algorithms. The color-color cut methods consist of ultraviolet (UV)-excess (e.g. \citealt{2009ApJS..180...67R}), the KX-technique (e.g. \citealt{2009A&A...494..579N, 2012MNRAS.424.2876M}), color selection of UV and optical data (e.g. \citealt{2011ApJ...728...23W}), color-color diagram based on optical and infrared data (e.g. \citealt{2010MNRAS.406.1583W,2016ApJ...819...24W}). As for variability selection, \citet{2011ApJ...728...26M} and \citet{2011AJ....141...93B} selected quasar candidates based on photometric variability by a damped random walk model;
\citet{2016A&A...587A..41P} presented the variability selection of quasars in eBOSS; \citet{2014MNRAS.439..703G} introduced Slepian wavelet variance (SWV) for quasar selection.

With the increase of astronomical data, machine learning algorithms are popular and widely applied on quasar candidate selection, for example, the probabilistic principal surfaces and the negative entropy clustering \citep{2009MNRAS.396..223D}, Bayesian selection method (e.g.,  \citealt{2004ApJS..155..257R,2011ApJ...743..125K,2015ApJ...811...95P}), artificial neural networks (ANNs; e.g. \citealt{2010A&A...523A..14Y,2015MNRAS.449.2818T}), extreme deconvolution (e.g. \citealt{2011ApJ...729..141B,2012ApJS..199....3R,2015ApJS..221...27M}), support vector machine (SVM; e.g. \citealt{2008MNRAS.386.1417G,2012MNRAS.425.2599P}), random forest algorithm (e.g. \citealt{2017ApJ...851...13S}). Other techniques based on radio, X-ray bands are also developed. \citet{2009AJ....138.1925M} obtained quasar candidates using radio criteria from SDSS database; \citet{2016AstL...42..277K} selected quasar candidates among X-Ray sources from the 3XMM-DR4 survey of the XMM-Newton observatory.

Besides SDSS, Pan-STARRS is another important optical sky survey project.
Therefore it is of great value for increasing the number of quasar candidates from Pan-STARRS.
Based on Pan-STARRS and WISE data, we aim at the efficient selection of quasar candidates in this paper.
The photometric data used for our candidate selection as well as known quasars and stars are presented,
and the rejection of extended objects is discussed (Section 2). Section 3 describes the color-color cut
to perform the selection of quasar candidates based on near-IR/infrared photometry. The principles
of support vector machine (SVM) and XGBoost are introduced in Section 4. Machine-learning algorithms
are further explored to classify quasar candidates and their performances are compared.
In Section~5, an XGBoost regressor is constructed to estimate quasar redshifts and applied to our newly predicted quasar candidates.
At last, a quasar candidate catalogue with the estimated redshifts is obtained as well as another catalogue that meets the observed conditions of the LAMOST telescope.
The experimental results and their applications are finally summarized in Section~6.

\section{The Data}
\subsection{PS1 Photometry}

Pan-STARRS is a wide-field optical/near-IR survey telescope system located at the Haleakala Observatory on the island of Maui in Hawaii.
Similar to SDSS, it provides five band photometries ($g_{\mathrm {P1}},r_{\mathrm {P1}},i_{\mathrm {P1}},z_{\mathrm {P1}},y_{\mathrm {P1}}$). The PS1 survey \citep{2016arXiv161205560C} is the first part of Pan-STARRS to be completed.
The 3$\pi$ survey is the largest survey PS1 has performed, covering the entire north to $-30$ deg declination.
The 5$\sigma$ median limiting AB magnitudes in the five PS1 bands $grizy$ are 23.2, 23.0, 22.7, 22.1 and 21.1 mag, respectively.
The Pan-STARRS photometry is extinction-corrected by the extinction coefficients
$\alpha_{g}$, $\alpha_{r}$, $\alpha_{i}$, $\alpha_{z}$, $\alpha_{y}$ = 3.172, 2.271, 1.682, 1.322, 1.087 with the extinction values from the SDSS photometry \citep{2011ApJ...737..103S}.
We set \texttt{MagErr} $<$ 0.2 for $g_{\mathrm {P1}},r_{\mathrm {P1}},i_{\mathrm {P1}},z_{\mathrm {P1}},y_{\mathrm {P1}}$ bands to exclude sources with large uncertainty, and set \texttt{psf\_qf\_perfect} $>$ 0.95 to remove observations which land on bad parts of the detector according to the work \citep{2016ApJ...817...73H}.

\subsection{WISE Photometry}

WISE \citep{2010AJ....140.1868W} is an infrared-wavelength astronomical space telescope launched by NASA and has mapped the entire sky at 3.4, 4.6, 12 and 22 $\mu m$ ($W1$, $W2$, $W3$, $W4$).
The 5$\sigma$ limiting AB magnitudes of the AllWISE catalog in $W1$, $W2$, $W3$ and $W4$ bands are 19.6, 19.3, 16.7, and 14.6 mag, separately.
The angular resolutions in each band are 6.1, 6.4, 6.5 and 12 arcsec, respectively. We only use photometric data in $W1$ and $W2$ bands, because $W3$ and $W4$ are much shallower and have larger magnitude errors.

To remove poor quality sources, some restrictions are set on primary detection and information flags. Unknown sources that meet the following conditions will be removed:

(i) Sources with high magnitude error and low signal-to-noise ratio ($\sigma_{\mathrm W1}>0.3$, $\sigma_{\mathrm W2}>0.3$, $SNR_{\mathrm W1}<3$, $SNR_{\mathrm W2}<3$).

(ii) Sources sources that are marked as saturated in the detection ($\mathtt{W1sat}\neq0$, $\mathtt{W2sat}\neq0$).

(iii) Objects affected by diffraction spikes, ghosts, latent images, and scattered light in both $W1$ and $W2$ ($\mathtt{cc\_flags}\neq0$).

(iv) Marked as extended sources in WISE detection ($\mathtt{ext\_flg}\neq0$).

(v) Small-separation and same-Tile (SSST) detection ($\mathtt{rel}\neq0$).

(vi) Affected by the nearby sources in the process of profile-fitting ($\mathtt{nb}>1$).

We cross match the Pan-STARRS1 MeanObject table with AllWISE to obtain the source containing optical and infrared photometric information. The match radius is set to 3 arcsec.

\subsection{Spectroscopically Identified Quasar and Star samples}

We use the latest SDSS Data Release 14 Quasar Catalog (DR14Q; \citealt{2018A&A...613A..51P}) as the known quasar sample. It contains 526,356 spectroscopically identified quasars, among which 144,046 are new discoveries from the latest SDSS-IV eBOSS survey \citep{2015ApJS..221...27M,2016AJ....151...44D}. DR14Q also includes previous spectroscopically identified quasars from SDSS-I, II and III.

Spectroscopically identified stellar sample is acquired from the SpecPhotoAll table in SDSS DR14 using CASJOB. Similar to \citet{2012MNRAS.425.2599P}, the sample meets the limitation of \texttt{class} $=$ \texttt{STAR}, \texttt{sciencePrimary} $=$ 1, \texttt{Mode} $=$ 1 and \texttt{zWarrning} $=$ 0. The records with fatal errors
are rejected using \texttt{flags} such as \texttt{BRIGHT}, \texttt{SATURATED}, \texttt{EDGE} and
\texttt{BLENDED}. Finally, a catalogue of 747,852 spectroscopically identified stars is obtained through the query.

In this article, we focus on Pan-STARRS and AllWISE datasets, so we match the SDSS known quasar and star samples with Pan-STARRS and AllWISE to obtain PS1 and AllWISE photometry.
The match radius is set to 3 arcsec. After matching and adding photometric quality limits (See $\S$2.1 and $\S$2.2) to remove poor quality sources, we obtain our known samples with Pan-STARRS and WISE photometry. After that, the number of stars is 355,255, that of quasars is 261,735. Table~1 details the star sample that has all values of PS1 photometry and W1W2 photometry for each subclass.
The top panel of Figure~1 shows the $r$ magnitude distribution of quasars and stars. It can be seen that stars are much brighter than quasars in our known samples.
The median $r$ magnitude of stars is 17.2 while that of quasars is 19.8.
The bottom panel of Figure~1 shows the redshift distribution of known quasars.

\begin{table}
 \caption{Spectroscopic Classes of Known Stars.}
 \label{tab:anysymbols}
 \begin{tabular*}{\columnwidth}{@{}l@{\extracolsep{\fill}}c}
  \hline
  Spectral Class  &  No. of objects\\
  \hline
   O  & 105\\
   OB & 21\\
   B  & 302\\
   A  & 19,340\\
   F  & 143,882\\
   G  & 27,538\\
   K  & 85,001\\
   M  & 76,734\\
   L  & 148\\
   T  & 49\\
   WD & 848\\
   CV & 716\\
   Carbon & 571\\
   Total No. & 355,255\\
  \hline
 \end{tabular*}
\end{table}

\begin{figure}
\includegraphics[width=\columnwidth]{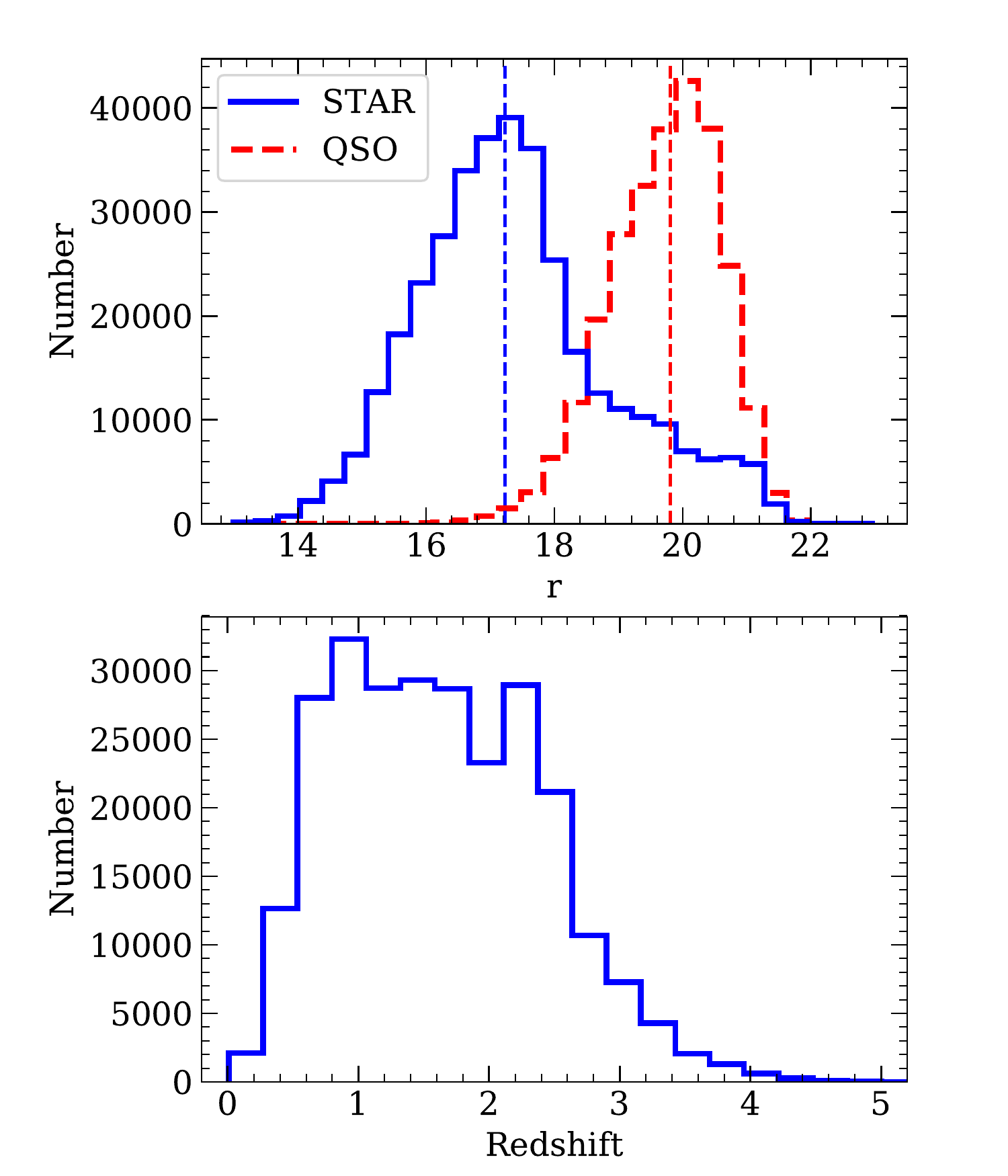}
\caption{The top panel shows the $r$ magnitude distribution of known quasars and stars that have all photometries of $g_{\mathrm P1},r_{\mathrm P1},i_{\mathrm P1},z_{\mathrm P1},y_{\mathrm P1},W1,W2$. The vertical dashed lines represent the median magnitude values of stars and quasars, respectively. The bottom panel shows the redshift distribution of known quasars that have all photometries of $g_{\mathrm P1},r_{\mathrm P1},i_{\mathrm P1},z_{\mathrm P1},y_{\mathrm P1},W1,W2$.}
\label{1}
\end{figure}

\section{Selection quasar candidates base on Optical and Infrared photometry}

\subsection{Discrimination of resolved and unresolved sources }

We focus on how to separate quasars from stars, so galaxies should be excluded. We use \texttt{PSFMag}$-$\texttt{KronMag} cut on $i$ and $z$ bands to exclude galaxies.
As shown in Figure~2, galaxies are clearly distinguished from stars and quasars in the distribution of \texttt{iPSFMag}$-$\texttt{iKronMag} and \texttt{zPSFMag}$-$\texttt{zKronMag}. The cut is set at \texttt{iPSFMag}$-$\texttt{iKronMag} $=$ 0.3 and \texttt{zPSFMag}$-$\texttt{zKronMag} $=$ 0.3 which is a strict limitation to exclude most galaxies. It can rule out $96.90\%$ of galaxies, with very few exceptions in our test on the known samples.
But it should be noted that this strategy would not exclude optically compact, unresolved sources that are selected by WISE to have a low-level active black hole but that do not have a luminous black hole in the optical band. Some of these might be considered as low-luminosity AGN rather than genuine quasars. Some of them might also be Type-1 quasars rather than having broad lines \citep{2003AJ....126.2125Z,2008AJ....136.2373R}.

Applying this criterion to the unknown sample obtained by crossing Pan-STARRS1 and AllWISE, we obtain the pointed sources to be predicted in the following.
\begin{figure}
\includegraphics[width=\columnwidth]{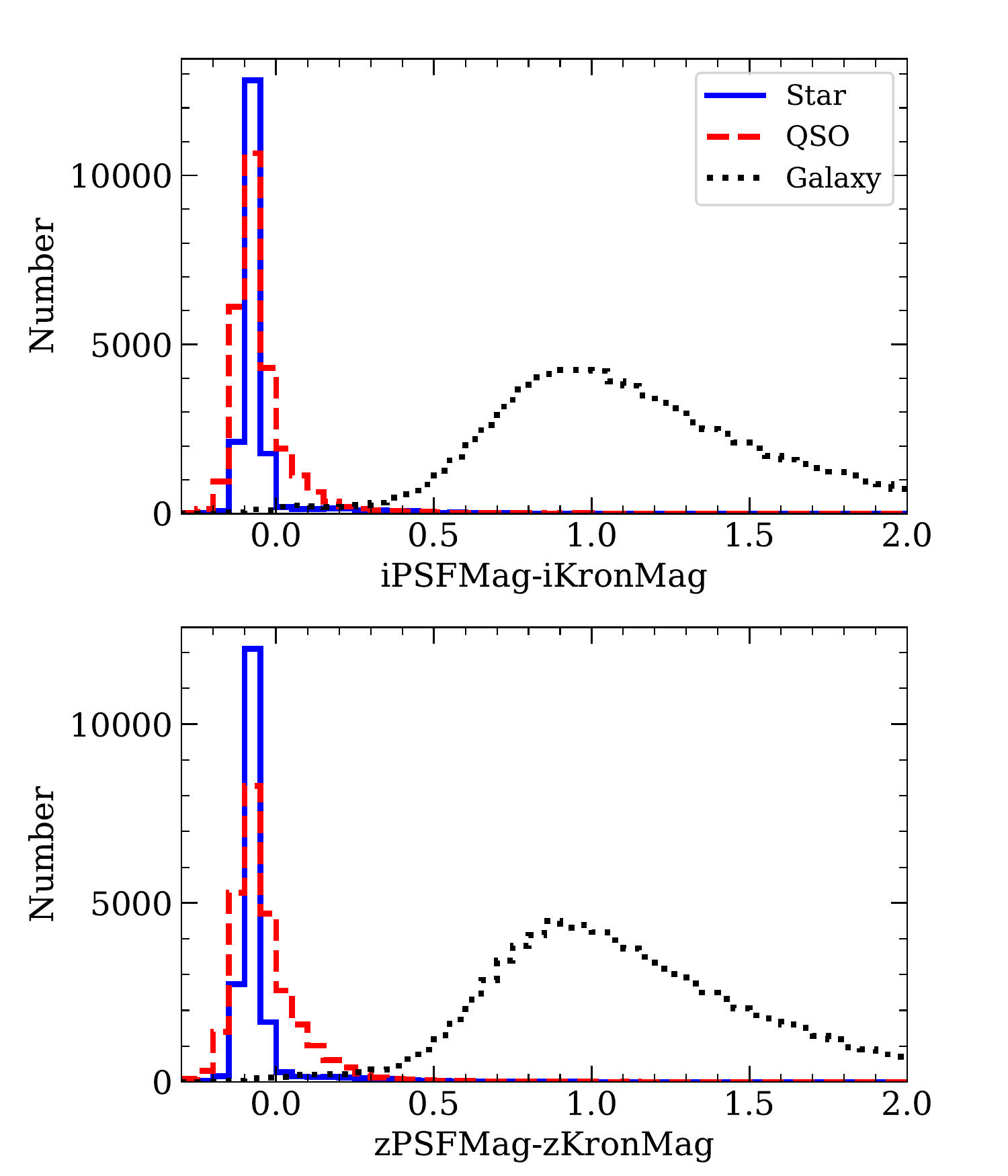}
\caption{The distribution of \texttt{iPSFMag}$-$\texttt{iKronMag} and \texttt{zPSFMag}$-$\texttt{zKronMag} for known stars, quasars and galaxies.}
\label{1}
\end{figure}

\subsection{The optical and infrared color cuts}

We aim to find quasar candidates through more bands of photometric information from both optical and infrared bands. For this purpose, we describe known quasars and stars in the color-color space. In Figure~3 we give six color-color diagrams. The quasars are indicated in different colors according to their redshifts.

\begin{figure*}
\includegraphics[width=\textwidth]{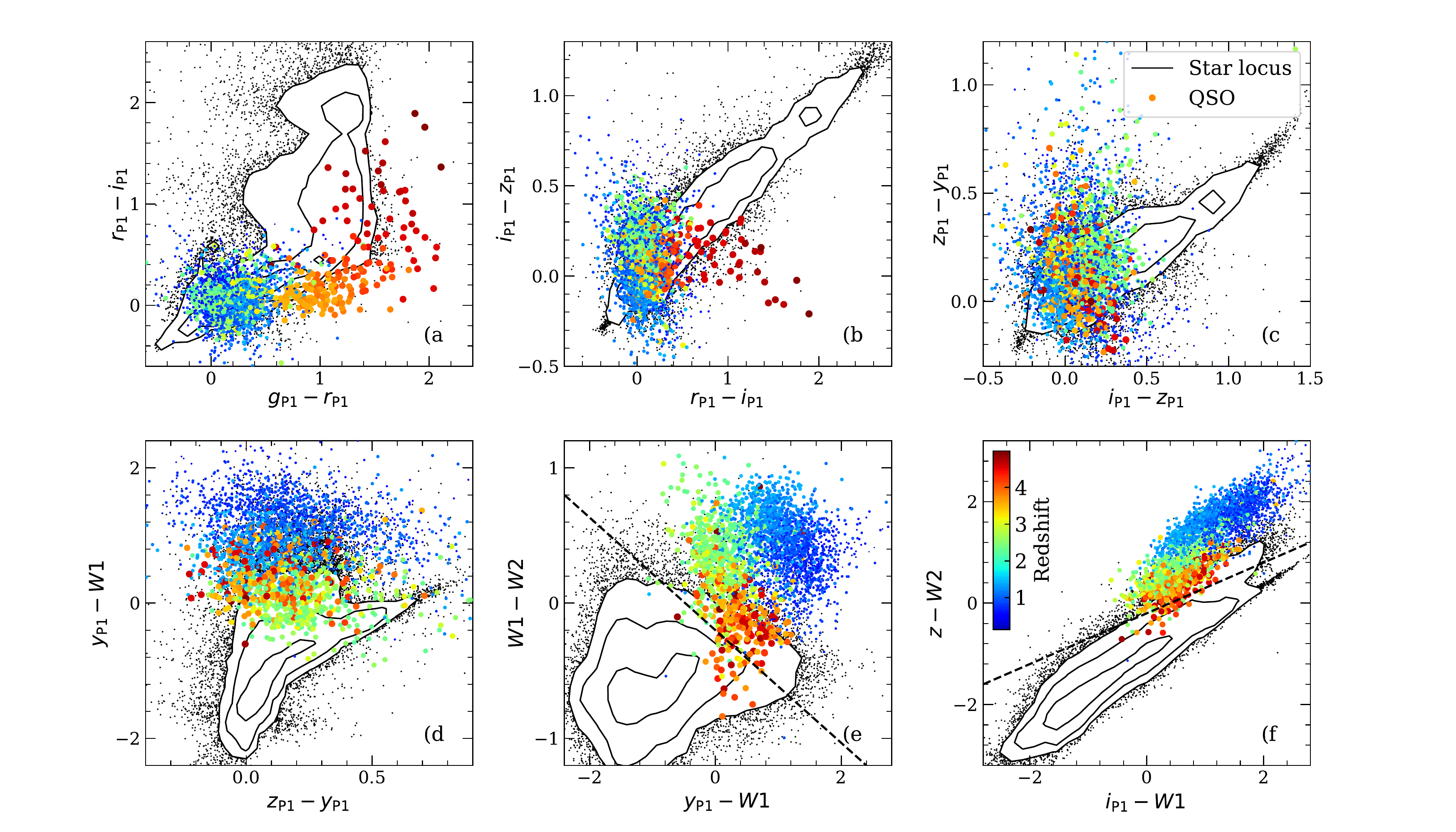}
\caption{The distribution of known stars and quasars in the color spaces of Pan-STARRS1 and AllWISE photometry.
The black outlines are the contours of star distribution and contain 68\%, 94\%, and 98.5\% of the stars, respectively.
Quasars are drawn in different colors and sizes according to their redshift values.
In the panels~(a)-(c), stars and quasars mostly overlap, in panel~d, they overlap partly, while in panels~(e)-(f) they are clearly separated.
The new color cuts are displayed in dashed line.}
\end{figure*}

It is easily found that quasars and stars are clearly separated in some color-color spaces (e.g. panels~e-f), while more overlapping in other panels (e.g. panels~a-d).
It indicates that adding infrared photometry is very effective for selecting quasar candidates, as other authors have found \citep{2012ApJ...753...30S,2013ApJ...772...26A}, which overcomes the inefficiency of the pure optical selection.
In the color space $y-W1$ vs. $W1-W2$ (panel~e), quasars and stars occupy different regions, with few overlap.
After examining known samples, we determine the cut that best separates the quasars and stars. In AB magnitude, the \texttt{yW1W2} cut can be written as
\begin{equation}
%\begin{aligned}
W1-W2 \geq -0.2 -0.417\times(y_{\mathrm P1}-W1)\\
%\end{aligned}
\end{equation}
In the test, the above color criterion produces a very pure quasar sample from the mixture of stars and quasars. Of all 355,255 stars, 348,530($98.11\%$) of stars are ruled out, while 6,725 escape the color criterion. Of all 261,735 quasars, 260,095($99.37\%$) are included in the color criterion. However, by displaying quasars in different colors according to their redshift values, we find that higher redshift quasars are more likely to mix with stars.

In order to test the effect of this color criterion for selecting different redshift quasars, we divide the quasars into two parts: high redshift ($z>3.5$) and low redshift ($z\leq3.5$). The result is that $11.38\%$ (424/3,724) high redshift quasars have been classified as stars while for low redshift quasars the ratio is only $0.47\%$ (1,216/258,011). Consistent with what Panel~(e) indicates, this color criterion is more effective for selecting low redshift ($z\leq3.5$) quasar candidates.

Another color-color space that is effective for distinguishing quasars from stars is the $i_{\mathrm P1}-W1$ vs. $z_{\mathrm P1}-W2$ color space (Panel~(f)), in which quasars and stars are located in two different areas.
After examining known samples, we obtain the \texttt{iW1zW2} cut in AB magnitude that clearly discriminates quasars from stars.
\begin{equation}
%\begin{aligned}
 z_{\mathrm P1}-W2 \geq -0.2 + 0.5\times(i_{\mathrm P1}-W1)
%\end{aligned}
\end{equation}
This color criterion also greatly reduces the pollution of stars, with 349,217($98.30\%$) of stars removed, while keeping 260,427($99.50\%$) of quasars. For $z>3.5$ quasars, $9.43\%$ (351/3,724) are mixed into stars, while for $z\leq3.5$ quasars, the value is $0.37\%$ (957/258,011).

In summary, the \texttt{yW1W2} and \texttt{iW1zW2} color cuts are very effective in selecting quasar candidates from the mixture of stars and quasars. Using these two color cuts, the majority of star contamination will be excluded, while the majority of quasars will be retained. However, one of the weaknesses of these two color cuts is that the high redshift quasars have a greater chance of being removed than the low redshift quasars. It seems difficult to completely exclude stars while retaining the vast majority of quasars only in a two-dimensional color-color space. So in this paper, we aim to use machine learning to select quasar candidates, in which stars and quasars can be better separated in higher dimensional color spaces.

\section{Selection Quasar Candidates Base on Machine learning}

\subsection{Introduction to SVM}

Support Vector Machines (SVM; \citealt{Vapnik1995The}) is a supervised machine learning algorithm used to solve classification and regression problems, especially for binary classification problems. The core idea of SVM is to find the best separation hyperplane in the feature space to separate the two classes. SVM maps the original feature vectors to a higher-dimensional space through the kernel function. Among all hyperplanes that can separate positive and negative samples, the best hyperplane is the one that has the largest margin between the data on both sides.

In binary classification, the training sample can be expressed as $T=\{\bmath x_{1},y_{1}\},...,\{\bmath x_{n},y_{n}\}$, where ${\bmath x}_{i} \in \mathtt{R}^{N}$, $y_{i}$ is the class label with the values $\{+1,-1\}$ for positive and negative classes, respectively. The separating hyperplane can be written as
\begin{equation}
\bmath{\omega}\cdot \bmath{x}_{i}+b=0\qquad    i=1,2,.....n\\
\end{equation}
We can get the optimal hyperplane with maximum margin by minimizing the following
\begin{equation}
F(\bmath{\omega},\bmath{\xi})=\frac{1}{2}(\bmath{\omega}\cdot\bmath{\omega})+C\sum_{i=1}^{n}\xi_{i}\\
\xi_{i}>0\\
\end{equation}
subject to
\begin{equation}
y_{i}[(\bmath{\omega}\cdot \bmath{x}_{i}+b)]\geq1-\xi_{i}\qquad  i=1,2,.....n\\
\end{equation}
Here $\xi$ is the slack variable which is introduced to allow the presence of points that are not completely separated by the hyperplane. The factor $C$ is the corresponding penalty coefficient for misclassification. The problem above can be solved by introducing the Lagrangian function, and it becomes the problem of solving the Lagrange multipliers $\alpha_{i}$ and $b$, as follows:
\begin{equation}
L(\omega,b,\alpha)=\sum_{i=1}^{n}\alpha_{i}-\frac{1}{2}
\sum_{i,j=1}^{n}\alpha_{i}\alpha_{j}y_{i}y_{j}x_{j}x_{i}^{T}\\
\end{equation}
subject to
\begin{equation}
\begin{split}
&\alpha_{i}[1-y_{i}(\sum_{i=1}^{n}\alpha_{i}y_{i}\langle x_{j},x_{i}\rangle+b)]=0\\
&\sum_{i=1}^{n}\alpha_{i}y_{i}=0\\
&0\leq\alpha_{i}\leq C\\
\end{split}
\end{equation}

SVM has been widely used in astronomy for star and galaxy classification \citep{2003PASP..115.1006Z,2004A&A...422.1113Z,2014NewA...28...35B,2018Ap&SS.363..140L,2018MNRAS.474.5232S}, quasar candidate selection \citep{2008MNRAS.386.1417G,2011ASPC..442..123Z,2011ASPC..442..447K,2012MNRAS.425.2599P} as well as quasar redshift estimation \citep{2008ASPC..394..509W,2016RAA....16...74H}.

\subsection{Introduction to XGBoost}

XGBoost \citep{Chen2016XGBoost} is a boosting algorithm that can be used for classification and regression problems.
It was developed on the basis of Gradient Boosting Decision Tree (GBDT; \citealt{Friedman2001Greedy}). Both XGBoost and GBDT are implementation of boosting method which aims to build strong classifiers by learning multiple weak classifiers.
The biggest difference between XGBoost and GBDT is that GBDT only uses the first derivative of the loss function to calculate the residual, while XGBoost applies not only the first derivative but also the second derivative of the loss function.
The predictors XGBoost builds are regression tree ensemble. For a given data set with $n$ examples and $m$ features $\texttt{D}=\{(x_{i},y_{i})\} (x_{i}\in \texttt{R}, y_{i}\in \texttt{R})$.
Set the model has $k$ trees totally. The prediction on $x_{i}$ is given by
\begin{equation}
\widehat{y_{i}} = \sum_{k=1}^K {f_{k}}(x_{i}),\quad {f_{k}}\in \texttt{F}
\end{equation}
where $f_{k}$ is one regression tree, and $f_{k}(x_{i})$ is the score it gives to $x_{i}$.
$\texttt{F}=\{f(x)=\omega_{q(x)}\}(q:\texttt{R}\rightarrow T,\omega\in \mathtt{R}^{\mathtt{T}})$ is the space of regression trees. Here $q(x)$ represents the structure function which is used to map each point to a leaf index.
The learning of the set of functions used in the model is done by minimizing the following regularized objective:
\begin{equation}
Obj = \sum_{i}^{n} l(\widehat{y_{i}},y_{i}) + \sum\Omega(f_{k})\\
\end{equation}
where the first term $l$ is the loss function and the second term $\Omega$ penalizes the complexity of the model.
We first discuss the first item. At the $t$-th iteration, it can be written as
\begin{equation}
\begin{aligned}
Obj^{(t)} &= \sum_{i}^n l(y_{i},\widehat{y_{i}}^{(t)}) + \Omega(f_{t})\\
          &= \sum_{i}^n l(y_{i},\widehat{y_{i}}^{(t-1)} + f_{t}(x_{i})) + \Omega(f_{t})
\end{aligned}
\end{equation}
The loss function can be written in the form of Taylor expansion and keep second-order accuracy.

\begin{equation}
Obj^{(t)} = \sum_{i=1}^N [g_{i}f_{t}(x_{i}) + \frac{1}{2} h_{i}f_{t}^{2}(x_{i})] + \Omega(f_{t})
\end{equation}
The $\Omega$ term is defined as complexity of the tree, it can be written as
\begin{equation}
 \Omega(f_{t}) = \gamma T + \frac{1}{2}\lambda\sum_{k=1}^{T}\omega_{k}^{2}
\end{equation}
Here $T$ represents the number of leaves, $\omega_{j}$ represents the score given by the $j$-th
leaf.
Define $I_{j} = \{i|q(x_{i})=j\}$ as the instance set of leaf $j$. The objective is then expressed as
\begin{equation}
 Obj^{(t)} = \sum_{j=1}^{T} [(\sum_{i\in I_{j}} g_{i})\omega_{j} + \frac{1}{2}(\lambda + \sum\limits_{i\in I_{j}} h_{i})\omega_j^{2}] + \gamma T
\end{equation}
If the structure of a tree $q(x)$ is fixed, we can gain the optimal weight of leaf $j$ by setting the derivative of $\omega_{j}$ equal to zero. Setting $G_{i}=\sum_{i\in I_{j}}g_{i}$ and $H_{i}=\sum_{i\in I_{j}}h_{i}$, $\omega_{j}$ is changed as following
\begin{equation}
\omega_{j}^{*} = -\frac{G_{i}}{H_{i}+\lambda}
\end{equation}
So $Obj^{(t)}$ can be simplified as
\begin{equation}
Obj^{(t)}(q) = -\frac{1}{2}\sum_{j=1}^{T} \frac{G_{i}^{2}}{H_{i}+\lambda} + \gamma T
\end{equation}
In practice, an optimized greedy algorithm is always used to grow the tree. It starts from a single leaf and tries to add a best spilt for each existed leaf node, iteratively. After adding a spilt, the objective turns as follows:
\begin{equation}
\mathrm{Gain} = \frac{1}{2}[-\frac{(G_{L}+G_{R})^{2}}{H_{L}+H_{R}+\lambda}+\frac{(G_{L}^{2})}{H_{L}+\lambda}+\frac{G_{R}^{2}}{H_{R}+\lambda}]-\gamma
\end{equation}
Gain is used to find the best spilt. For one possible spilt, XGBoost scans from left to right of the leaves to calculate the $G_{L}$ and $G_{R}$, $H_{L}$ and $H_{R}$, then the Gain can be calculated. Note that the formula has a penalty term $\gamma$ which can lead to a negative gain. A negative gain means that the newly added spilt does not make the algorithm perform better and should be cut off.

In astronomy, XGBoost was recently used for the separation of pulsar signals from noise \citep{2018A&C....23...15B} and the classification of unknown source in the Fermi-LAT catalog \citep{2016ApJ...825...69M}.

\subsection{Classification Metrics}
There are many criterions to score the performance of a classifier. Here we only apply three standard metrics (\textrm{Accuracy}, \textrm{Precision} and \textrm{Recall}) to determine which one classifier is better. In binary classification, the accuracy is the ratio of the number of the correctly classified samples to the total number of the two classes of samples.
\begin{equation}
 \mathrm{Accuracy} = \frac{TP+TN}{TP + TN + FP + FN}
\end{equation}
Here TP is the positive sample that is correctly classified by the classifier. FP represents the negative sample that is classified as positive. FN refers to the positive sample that is classified as negative.

The precision (P) is the ratio of the true positive (negative) sample in all the samples that are predicted to be positive (negative).
The Recall (R) is the ratio of correctly predicted positive (negative) samples in all the true positive (negative) samples.
\begin{equation}
 \mathrm{Precison} = \frac{TP}{TP + FP}, \mathrm{Recall} = \frac{TP}{TP + FN}
\end{equation}

\subsection{Binary Classifiers Construction of XGBoost and SVM}
In this section, we detail how to use photometric information to construct XGBoost and SVM classifiers for quasar selection.
The quasar and star training sets are introduced in \S2.3. The input pattern used for SVM and XGBoost training is a combination of colors.
To test the effect of different input patterns on the performance of a classifier, we build a classifier using different input patterns.
The input feature is the colors, which is obtained from the magnitude difference between two different bands, like ($g_{\mathrm P1}-r_{\mathrm P1}$,$r_{\mathrm P1}-i_{\mathrm P1}$,$i_{\mathrm P1}-z_{\mathrm P1}$), and so on, where the magnitudes have been corrected for dust extinction through our galaxy.

All the input patterns we used are detailed in Table~2. We use the XGboost and SVM python packages provided by scikit-learn \citep{scikit-learn}. For short, the default model parameters are adopted to test the validity of different input features in the first step,
and then the optimal model will be gained by a grid search on the hyperparameters.
For all of these experiments, we use 5-fold cross-validation to train and test the classifier, which means that the classifier will always be trained on 80\% of the entire training set and tested on the the remaining 20\% of the set.
The cross-validation scores of XGBoost and SVM classifiers are given in Table~2.

As shown in Table~2, the XGBoost and SVM classifiers both achieve very high scores when distinguishing stars and quasars.
The performance of XGBoost is better than or comparable to that of SVM. When pure optical information is considered, XGBoost scores apparently higher than SVM. Given mixed infrared and optical information, XGBoost and SVM yield similar scores. However, in terms of the computing time, XGBoost is at speed far faster than SVM when only one CPU is used. At present, the program of XGBoost supports parallel computation while that of SVM doesn't support. If adopting parallel computation, the speed of XGBoost will be more faster. For short, we write the total color combination as \texttt{8Color}, which is $g_{\mathrm P1}-r_{\mathrm P1},r_{\mathrm P1}-i_{\mathrm P1},i_{\mathrm P1}-z_{\mathrm P1},z_{\mathrm P1}-y_{\mathrm P1},y_{\mathrm P1}-W1,W1-W2,i_{\mathrm P1}-W1,z_{\mathrm P1}-W2,r_{\mathrm P1}$.
Table~2 also indicates that different input patterns have great impact on the performance of a classifier, moreover the combination of optical and infrared bands can achieve higher scores than pure optical band. In other words, considering the classification criterions (\textrm{Accuracy}, \textrm{Precision} and \textrm{Recall}), the performance increases when adding information from infrared band for XGBoost and SVM. The best performance with \texttt{8Color} is achieved given information from both optical and infrared bands. As a result, \texttt{8Color} should be considered as the input pattern in the training to improve classification and regression accuracy in the following experiments. In brief, adding information from infrared band is helpful to select out more quasar candidates \citep{2012ApJ...753...30S,2013ApJ...772...26A}.

\begin{table*}
\caption{The performance of XGBoost and SVM classifier with different input features}
\label{tab:anysymbols}
\tiny{
\begin{tabular*}{\textwidth}{@{}l@{\extracolsep{\fill}}ccccccc}
  \hline
           Features & Algorithm  &$\mathrm{Accuracy}$(\%) & $\mathrm{Precision^{+}}$(\%) & $\mathrm{Recall^{+}}$(\%)  & $\mathrm{Precision^{-}}$(\%) & $\mathrm{Recall^{-}}$(\%)  & Time(s)\\
          \hline
         $g-r,r-i,i-z,z-y$   & XGBoost & 96.07 & 96.42 & 94.23 & 95.82 & 97.42 & 27.74$\pm$0.25 \\
         $i-z,z-y,y-W1,W1-W2$   & XGBoost & 99.24& 98.91 & 99.31 & 99.49 & 99.20 & 27.32$\pm$0.1\\
         $y-W1,W1-W2,i-W1,z-W2$   & XGBoost & 99.14 & 98.78 & 99.20 & 99.41 & 99.10 & 26.95$\pm$0.07 \\
         \texttt{8Color}               & XGBoost & 99.46 & 99.19 & 99.53 & 99.65 & 99.40 & 46.26$\pm$0.12 \\
        \hline
         $g-r,r-i,i-z,z-y$  & SVM  & 92.96 & 96.77 & 86.29 & 90.65 & 97.88 & 6,229$\pm$526 \\
         $i-z,z-y,y-W1,W1-W2$  & SVM & 99.22 & 98.80 & 99.37 & 99.53 & 99.11 & 431$\pm$46 \\
         $y-W1,W1-W2,i-W1,z-W2$  & SVM & 99.20 & 98.73 & 99.40 & 99.55 & 99.06 & 701$\pm$241 \\
         \texttt{8Color}  & SVM & 99.46 & 99.11 & 99.63 & 99.73 & 99.34 & 389$\pm$58  \\
        \hline
\multicolumn{8}{l}{$^a$ In the table, $g_{\mathrm P1},r_{\mathrm P1},i_{\mathrm P1},z_{\mathrm P1},y_{\mathrm P1}$ are written as $g,r,i,z,y$ for simplicity.}\\
\multicolumn{8}{l}{$^b$ \texttt{8Color} represents $g_{\mathrm P1}-r_{\mathrm P1},r_{\mathrm P1}-i_{\mathrm P1},i_{\mathrm P1}-z_{\mathrm P1},z_{\mathrm P1}-y_{\mathrm P1},y_{\mathrm P1}-W1,W1-W2,i_{\mathrm P1}-W1,z_{\mathrm P1}-W2.$}\\
\end{tabular*}}
\end{table*}

When the input pattern and training samples are specified, the performance of a classifier depends on its model parameters.
A core issue in machine learning is to avoid over-fitting, which means that the algorithm learns too detailed information on the training set, resulting in weak generalization of the model to other data sets. The opposite is the under-fitting, in which case the model learns too rough information.
Several model parameters in XGBoost are used to prevent the model from over-fitting and under-fitting.
For example, the parameter $max\_depth$ sets the maximum depth of the decision tree. As the value of $max\_depth$ increases, the model learns more specific and detailed structures, but too high a value can lead to over-fitting.
So the hyperparameters of the algorithm need to be adjusted to accommodate different training data.
Since XGBoost has several hyperparameters, we use a limited grid search for some key hyperparameters.
The total hyperparameters and their optimal values are listed in Table~3.

For the SVM classifier with \texttt{8Color} as the input pattern and radial basis function as kernel function,
the grid search is applied for hyperparameters $C$ and $\gamma$, and finally the optimal values $C=1$, $\gamma=0.01$ are obtained.

After the step of model parameter optimization, the classification performance improves in some degree for both XGBoost and SVM. The final optimized XGBoost and SVM classifiers will be used for the following quasar candidate selection.

\begin{table}
\caption{XGBoost hyperparameters used for parameter optimization through grid search and the performance of the model before and after parameter optimization.}
\label{tab:anysymbols}
\setlength{\extrarowheight}{.3em}
%\font\tenrm = cmr10 at 6pt
%\fontfamily{pcr}\selectfont
\begin{tabular*}{\columnwidth}{@{}l@{\extracolsep{\fill}}cc}
  \hline
  Hyperparameters & Default Values & Optimal Values \\
  \hline
  \selectfont
  $\texttt{objective}$ & binary logistic&binary logistic\\
  $\texttt{booster}$ & gbtree & gbtree\\
  $\texttt{n\_estimators}$ & 100 & 1000 \\
  $\texttt{learning\_rate}$ & 0.1 & 0.01 \\
  $\texttt{max\_depth}$ & 3 & 7 \\
  $\texttt{min\_child\_weight}$ & 1 & 3 \\
  $\texttt{gamma}$ & 0 & 0.1 \\
  $\texttt{subsample}$ & 1 & 0.6 \\
  $\texttt{colsample\_bytree}$ & 1 & 0.9 \\
  $\texttt{colsample\_bylevel}$ & 1 & 1\\
  $\texttt{reg\_alpha}$ & 0 & 0 \\
  $\texttt{reg\_lambda}$ & 1 & 1\\
  $\texttt{max\_delta\_step}$ & 0 & 0\\
  $\texttt{scale\_pos\_weight}$ & 1 &1\\
  $\texttt{base\_score}$ & 0.5 & 0.5\\
  \hline
  $\mathrm{Accuracy}$ & 99.46\% & 99.58\%\\
  $\mathrm{Precision^{+}}$ & 99.19\%& 99.34\%\\
  $\mathrm{Recall^{+}}$ & 99.53\%& 99.67\%\\
  $\mathrm{Precision^{-}}$ &99.65\% &99.76\% \\
  $\mathrm{Recall^{-}}$ &99.40\% & 99.51\%\\
  \hline
\end{tabular*}
\end{table}

\subsection{Test on different samples}
After obtaining the optimal model parameters, we further check the influence of different training or test samples in different magnitude ranges on the performance of a classifier.
Taking XGBoost for example, we implement four experiments: training and test samples both with $(r<18.5)$, training and test samples both with $(r\geq18.5)$, training sample with $(r<18.5)$ while test sample with $(r\geq18.5)$, training sample with $(r\geq18.5)$ while test sample with $(r<18.5)$. The experimental results are shown in Table~4, which indicates that the performance of training and test samples both in bright magnitude range $(r<18.5)$ is the best, nevertheless the performances of other situations are satisfactory with different criterions larger than 95\%. As a result, whether training and test samples are in the same magnitude ranges influences on the performance of a classifier to a certain extent.

\begin{table*}
 \caption{Classification Results of XGBoost Classifier Trained on Different Training Sets and Test on Different Test Sets.}
 \label{tab:anysymbols}
 \begin{tabular*}{\textwidth}{@{}l@{\extracolsep{\fill}}ccccccc}
  \hline
        Training set & Test set & $\mathrm{Accuracy}$(\%) & $\mathrm{Precision^{+}}$(\%) & $\mathrm{Recall^{+}}$(\%)  & $\mathrm{Precision^{-}}$(\%) & $\mathrm{Recall^{-}}$(\%)\\
   \hline
        214659$(r<18.5)$& 91997$(r<18.5)$ & 99.96 & 99.67 &99.83 & 99.99 & 99.97\\
        217288$(r\geq18.5)$& 93123$(r\geq18.5)$ & 99.33 & 99.42 &99.72 & 99.06 & 98.05\\
        306658$(r<18.5)$& 310413$(r\geq18.5)$ & 98.51 & 99.43 &98.63 & 95.57 & 98.10\\
        310413$(r\geq18.5)$& 306658$(r<18.5)$ & 99.93 & 99.14 &99.88 & 99.99 & 99.93\\
  \hline
 \end{tabular*}
\end{table*}

In order to verify the efficiency of our classifiers, we need choose a highly spectroscopically complete region of the SDSS to determine what fraction of our candidates to the SDSS/eBOSS spectroscopic limit (about $r\sim 21.5$) are quasars or stars. For this, we apply the region called ``SEQUELS" for such a comparison, which is described in \citep{2017A&A...597A..79P, 2015ApJS..221...27M}. SEQUELS includes two chunks of BOSS, located at $120^\circ \leq \alpha_{\mathrm J2000}<210^\circ$ and $+45^\circ \leq \delta_{\mathrm J2000}<+60^\circ$, and covers 810 deg$^2$ in total area. We perform our classifiers on the known star and quasar samples in this region, the results are indicated in Table~5. Table~5 shows that our classifiers are applicable with the recall more than 98.4\% for stars or quasars. In addition, we use our classifiers on the unknown sample in this region and obtain 52,685 quasar candidates with $P_{\mathrm {QSO}}>0.5$ and 21,795 quasar candidates with $P_{\mathrm {QSO}}>0.95$ among 12,447,801 unknown sources.

\begin{table}
\caption{The classification results in the SEQUELS region by XGBoost and SVM classifiers.}
 \label{tab:anysymbols}
 \begin{tabular*}{\columnwidth}{@{}l@{\extracolsep{\fill}}ccc}
 %\hline
  %      \multicolumn{3}{l}{Train on QSO($r<18$) and then test on QSO($R>20$) }\\
  \hline
        XGBoost & Pred. STAR & Pred. QSO& \textrm{Recall}\\
  \hline
         STAR & 37,760 & 283& 99.26\% \\
         QSO & 839 & 54,021& 98.47\%\\
  \hline
   SVM & Pred. STAR & Pred. QSO& \textrm{Recall}\\
  \hline
         STAR & 37,591 & 452&98.81\% \\
         QSO & 869 & 53,991&98.42\%\\
  \hline
 %\multicolumn{5}{l}{XGB$\cap$SVM means the intersectiont of XGBoost predict result and SVM predict result}\\
 \end{tabular*}
\end{table}

\subsection{Application of the Classifiers on Pan-STARRS1 and AllWISE Databases}

Up to now, lots of Pan-STARRS1 sources haven't been recognized from spectra. We aim to improve the efficiency of spectral recognition and discover more quasars, then devise a schema to select quasar candidates.
Firstly the unknown data are collected from Pan-STARRS1 and AllWISE cross-matched databases meeting the limitations described in Section~2, and then the constraints \texttt{rPSFMag}$-$\texttt{iKronMag} $=$ 0.3 and \texttt{iPSFMag}$-$\texttt{zKronMag} $=$ 0.3 are applied to rule out the extended sources. In order to obtain the quasar candidates as efficient as possible, \texttt{8Color} is adopted as the input pattern of XGBoost and SVM classifiers.
The optimal XGBoost and SVM classifiers are constructed in Section~4.4. We apply the classifiers on the unknown pointed sample from the matched Pan-STARRS and AllWISE datasets to find new quasar candidates.
These unknown sources are taken as the input of the XGBoost and SVM classifiers and the predicted results are output with their probabilities. Given the predicted probability $P>0.95$ of being a quasar, the XGBoost classifier obtains 1,299,304 new quasar candidates, while SVM classifier yields 1,365,239 new quasar candidates, and the sources assigned as quasar candidates by both of these two classifiers add up to 1,201,211. If obtaining more complete quasar candidates, the combined results of the two methods may be used. If getting more reliable quasar candidates, the cross-matched result of the two methods is better. The detailed distributions in different probability ranges for the predicted results are shown in Table~6.

In general, any model has its advantages and disadvantages; it is created from the training sample and then limited by the training sample. No matter for the test or unknown sample, the sample should be similar to the training sample in magnitude range when it is predicted by any created model. Therefore any model should be recreated for keeping its scalability and efficiency when more new known data are obtained.

\begin{table}
 \caption{The Number of the quasar candidates in different probabilities by XGBoost and SVM classifiers}
 \label{tab:anysymbols}
 \begin{tabular*}{\columnwidth}{@{}l@{\extracolsep{\fill}}cccc}
  \hline
        Method & Constraints & $P>0.5$  & $P>0.75$ & $P>0.95$\\
  \hline
         XGBoost& ... & 2,173,846 & 1,791,487 & 1,299,304\\
         SVM & ... & 2,215,756& 1,864,498 & 1,365,239\\
         XGB$\cap$SVM & ... & 2,006,632& 1,657,988 & 1,201,211\\
  \hline
         XGBoost& $r<20$ & 533,258 & 465,360 & 401,507\\
         SVM &$r<20$ & 582,002& 506,651 & 423,566\\
         XGB$\cap$SVM & $r<20$ & 506,768& 448,515 & 390,208\\
  \hline
 %\multicolumn{5}{l}{XGB$\cap$SVM means the intersectiont of XGBoost predict result and SVM predict result}\\
 \end{tabular*}
\end{table}

To verify the rationality of the predicted quasar candidates in different magnitude ranges, we examine the quasar luminosity function (QLF) based on different models, which was raised by \citet{2016A&A...587A..41P}.
They used the variability-selected quasars in the SDSS Stripe 82 region to provide a new measurement of the quasar luminosity function (QLF) in the redshift range of $0.68< z < 4.0$. They fitted the QLF using two independent double-power-law models, one was a pure luminosity-function evolution (PLE) and the other was a simple PLE combined with a model that comprises both luminosity and density evolution (LEDE).
They provided the predicted quasar counts according to the two QLFs for a survey covering 10,000 deg$^{2}$ (see their Table~7 in \citealt{2016A&A...589C...2P}).
According to their calculations, the PLE model predicted 1,144,614 quasars in over $15.5 < r < 21.5$ and $0 < z < 6$ for a survey covering 10,000 deg$^{2}$ while the PLE+LEDE model predicted 1,152,555 quasars in the same region.
The PS1 survey covers $3\pi$ steradian, including the entire northern sky and the $0\leq$ dec $\leq30$ southern sky, which is approximately equal to 30,940 square degrees. According to their model, there should be at least 3,000,000 quasars observed throughout the sky in $15.5 < r < 21.5$ observed in the PS1 survey. According to our predictions, a total of 1,945,729 sources were predicted both by XGBoost and SVM to be quasars within a range of 15.5 < r < 21.5, among them, 1,164,653 have high probability $P_{\mathrm {QSO}}>0.95$. It is apparent that our prediction of quasar candidates is in accordance with the model prediction.

\section{Photometric Redshift Estimation}

Since obtaining spectral redshifts is inefficient and expensive, using photometric information to obtain quasar redshifts has become a research interest \citep{2001AJ....122.1151R,2001AJ....122.1163B,2004ApJS..155..243W,2007ApJ...663..774B}.
\citet{2017AJ....154..269Y} presented a new algorithm to estimate quasar photometric redshifts (photo-zs) by considering the asymmetries in the relative flux distributions of quasars.
\citet{2017ApJ...851...13S} used SVM regression(SVR) and random forest (RF) regression to estimate quasar photometric redshifts with adjacent flux ratios.

For the new quasar candidates predicted both by XGBoost and SVM methods in Table~6, we aim to obtain their photometric redshifts for further screening.
Based on the same quasar training set as the classification (detailed description in the Section~1), we train the XGBoost regression and SVM regression to predict the quasar photometric redshifts with \texttt{8Color} as input pattern.

\subsection{Regression Metrics}

To evaluate the result of our photometric redshift estimation, the mean absolute error $\sigma$ and the $R^{2}$ scores are adopted.
The $\sigma$ is defined as
\begin{equation}
\sigma = \frac{1}{n}\sum_{i=0}^{n-1}|z_{i}-\widehat{z}_{i}|\\
\end{equation}
Here $z_{i}$ is the true redshift, $\widehat{z_{i}}$ is the the predicted redshift value and $n$ is the sample size.
The expression of $R^{2}$ is
\begin{equation}
R^{2}(z,\widehat{z})=1-\frac{\sum_{i=0}^{n-1}(z_{i}-\widehat{z}_{i})^{2}}{\sum_{i=0}^{n-1}(z_{i}-\overline{z}_{i})^{2}}\\
\end{equation}
In addition to these metrics, the fraction of test samples that satisfy $\mid\bigtriangleup z\mid=|z_{i}-\widehat{z_{i}}|<e$ is often used to evaluate the redshift estimation, where $e$ is the given threshold.
\begin{equation}
f_{\mid\bigtriangleup z\mid<e}=\frac{N(|z_{i}-\widehat{z_{i}}|<e)}{N_{total}}\\
\end{equation}
Usually the values of $e$ are 0.1, 0.2 and 0.3. However, the redshift normalized residuals are adopted in most cases.
\begin{equation}
\delta_{e}=\frac{N(|z_{i}-\widehat{z_{i}}|<e(1+z_{i}))}{N_{total}}\\
\end{equation}

\subsection{Regression Results and Their Application}

The photometric redshifts of the test quasar sample are estimated by XGBoost and SVM regression, comparing with spectral redshifts as shown in Figure~4. The scores for the regression metrics of the two methods are listed in the upper left corner in each panel of Figure~4. In order to further compare the performance of photometric redshift estimation for the two methods, the distribution of $\triangle z=z_{\mathrm {spec}}-z_{\mathrm {reg}}$ is indicated in Figure~5.
The XGBoost regression gives a slightly tighter histogram distribution around $\triangle z=0$. Thus with the same input pattern, XGBoost regression is a little superior to SVM regression at least to the flux limits of our training set.

\begin{figure}
\includegraphics[width=\columnwidth]{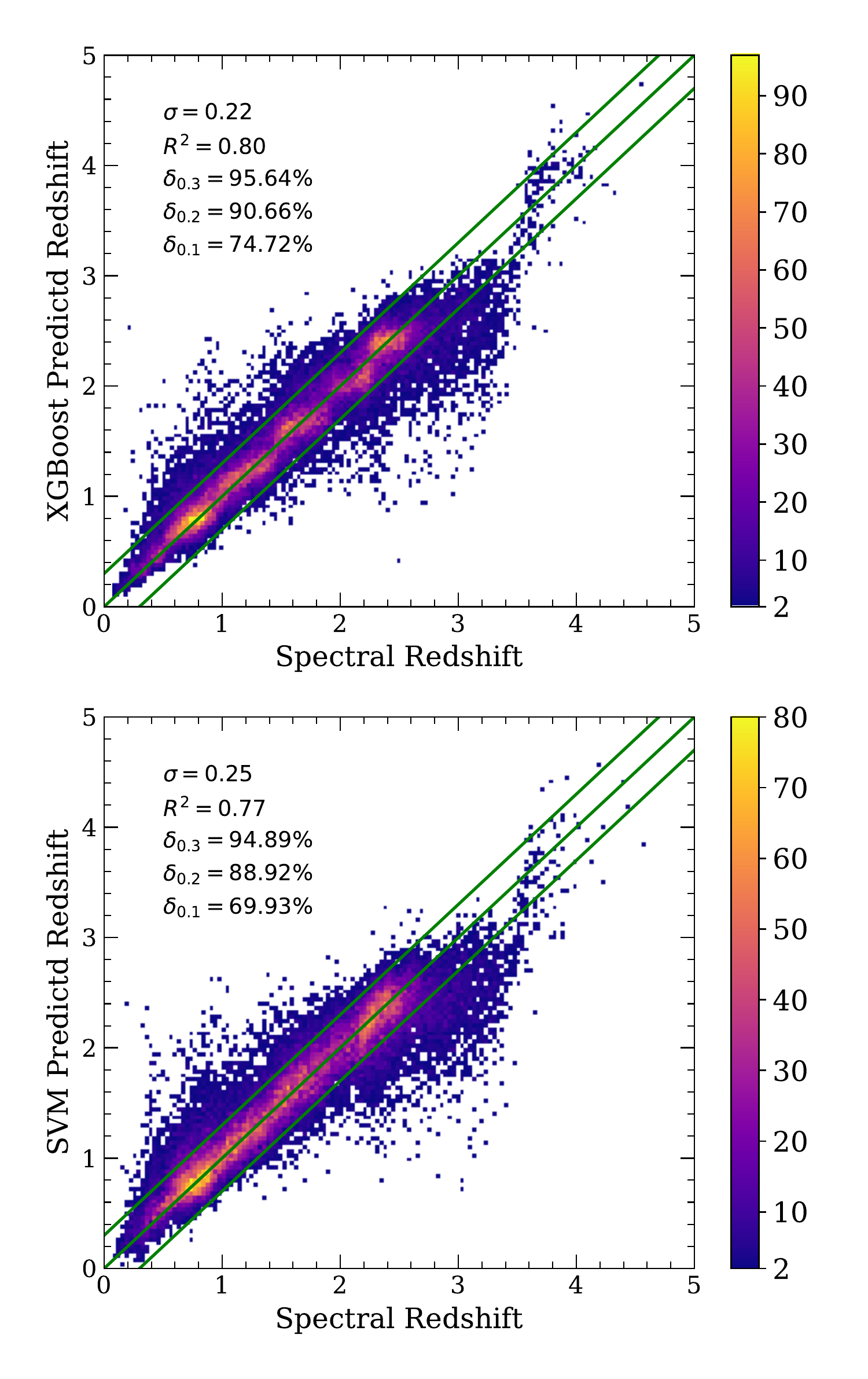}
\caption{The estimated redshift by XGBoost and SVM regression against spectral redshift on our test set with \texttt{8Color}. The color bar shows the density distribution and gives the number in each square bin. The three lines represent the diagonal of $\bigtriangleup z=0$ and the boundary of $\bigtriangleup z=0.3$}
\label{1}
\end{figure}

\begin{figure}
\includegraphics[width=\columnwidth]{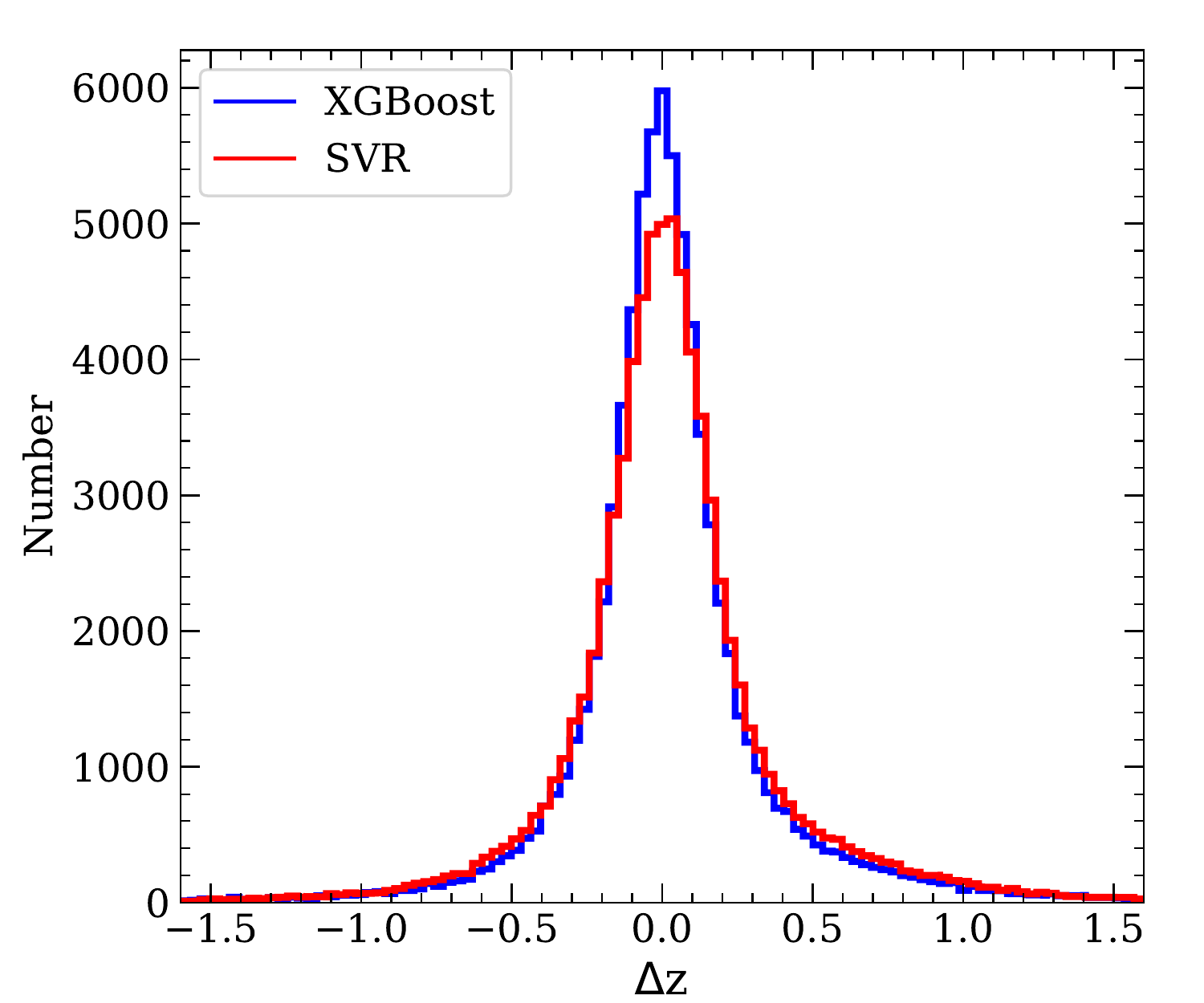}
\caption{Distribution of the difference between the spectroscopic redshifts and estimated redshifts $\triangle z=z_{\mathrm {spec}}-z_{\mathrm {reg}}$ for XGboost and SVM on the test quasar sample.}
\label{1}
\end{figure}

As a result, we choose the XGBoost method to estimate the photometric redshifts of the new quasar candidates with \texttt{8Color} as input pattern.
In order to investigate the redshift distribution of the new quasar candidates, we adopt Richards's criterion to divide quasars into three subgroups according to the reshift intervals $z\leq2.2$, $2.2<z\leq3.5$ and $3.5<z\leq5$ \citep{2009ApJS..180...67R}. Table~7 lists the count of quasar candidates in different probabilities and redshift intervals.
As shown in Table~7, only considering the probability $P_{\mathrm {QSO}}>0.95$, the number of low redshift, medium redshift and high redshift quasar candidates is 963,628, 230,181 and 7,402, respectively, when the magnitude $r<20$, the number of these candidates is separately 324,934, 62,674 and 2,600.
The number of different redshift quasar candidates is apparently different, of which the number of higher redshift quasars is comparatively smaller. This is consistent with the fact that higher redshift quasars are too much fainter to be observed. For simplification, we only probe the redshift distribution and space distribution of quasar candidates with $P_{\mathrm {QSO}}>0.95$ of intersected results predicted by XGBoost and SVM classifiers in Figures~6-7. Figure~6 describes the redshift distribution of different samples, i.e., spectroscopic redshift distribution of known SDSS quasar sample, photometric redshift distribution of the quasar candidates. It is found that the number of our newly predicted quasar candidates in the redshift range from 0.6 to 2.6 is much larger than that of known SDSS identified quasars while the candidates with high redshift $z>2.6$ are comparable to those of SDSS. Nevertheless, it is also noted here that the SDSS/BOSS spectroscopic survey covers about 10,000 deg$^2$, whereas our candidate sample covers about 30,000 deg$^2$. In order to learn the space distribution of quasar candidates, we plot the quasar candidates in the Galactic locations in Figure~7. As shown in Figure~7, the most of quasar candidates occupy the Galactic medium latitude and are far away from the Galactic plane while only small quantity of them focus on the Galactic plane.

\begin{table}
\caption{Quasar candidates in different redshift intervals}
\label{tab:anysymbols}
\begin{tabular*}{\columnwidth}{@{}l@{\extracolsep{\fill}}ccc}
\hline
&$P_{\mathrm {QSO}}>0.5$ & $P_{\mathrm {QSO}}>0.75$ & $P_{\mathrm {QSO}}>0.95$ \\
\hline
$z\leq2.2$ & 1,323,485 & 1,187,900 & 963,628 \\
$2.2<z\leq3.5$ & 617,536 & 436,811 & 230,181 \\
$z>3.5$ & 65,611 & 33,277 & 7,402 \\
\hline
Total No. & 2,006,632 & 1,657,988 & 1,201,211\\
 \hline
when $r<20$&&&\\
  \hline
$z\leq2.2$ & 342,912 & 337,198 & 324,934\\
$2.2<z\leq3.5$ & 147,354 & 102,963 & 62,674\\
$3.5<z\leq5$ & 16,502 & 8,354 & 2,600\\
\hline
Total No.& 506,768 & 448,515 & 390,208\\
\hline
\end{tabular*}
\end{table}

\begin{figure}
\includegraphics[width=\columnwidth]{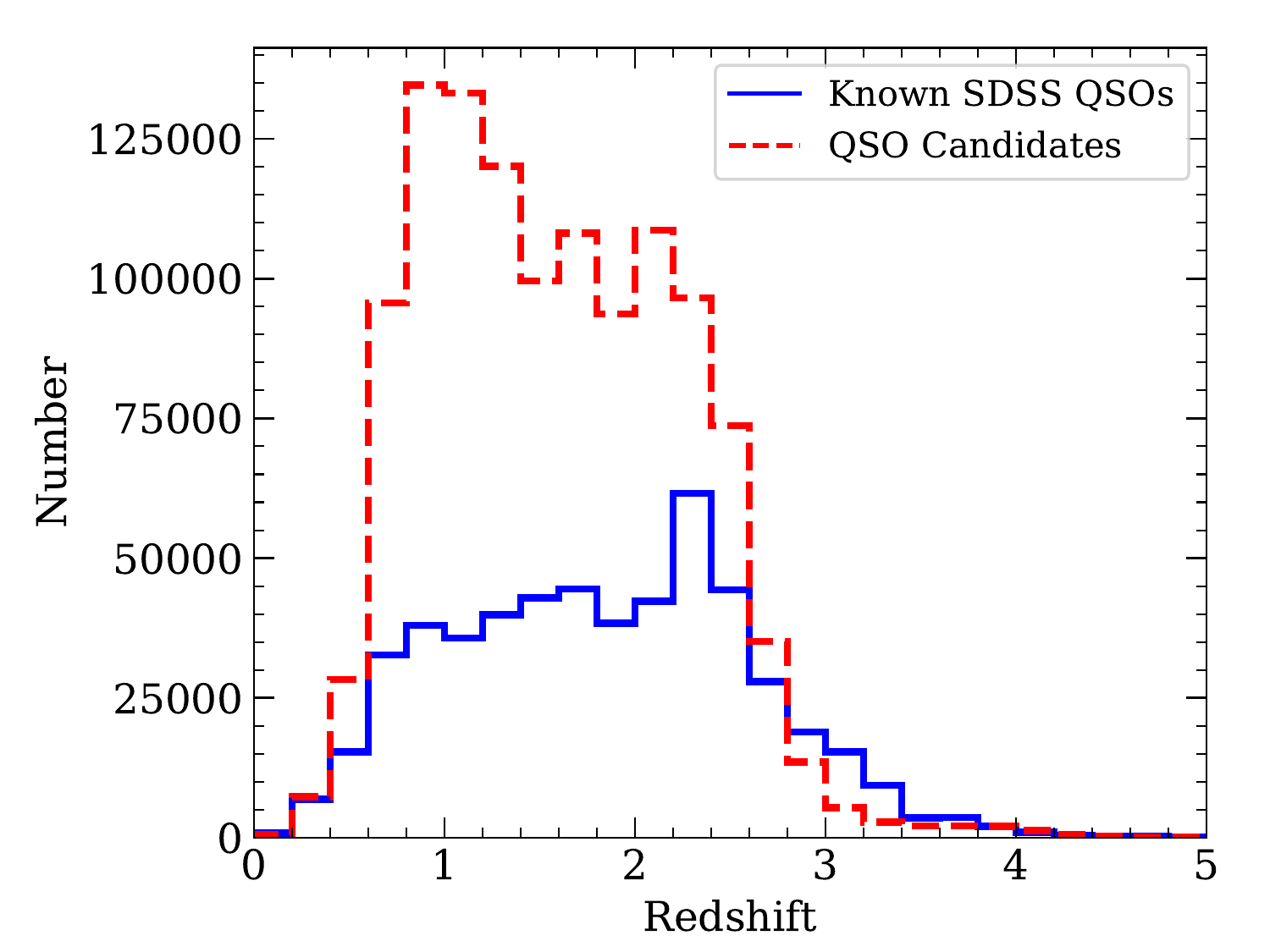}
\caption{Redshift distribution for different samples, solid line for the whole known SDSS quasars covering about 10,000 deg$^2$, dashed line for the newly predicted quasar candidates with $P_{\mathrm {QSO}}>0.95$ covering about 30,000 deg$^2$.}
\label{1}
\end{figure}

%bb=23 5 593 382,
\begin{figure}
\includegraphics[width=\columnwidth]{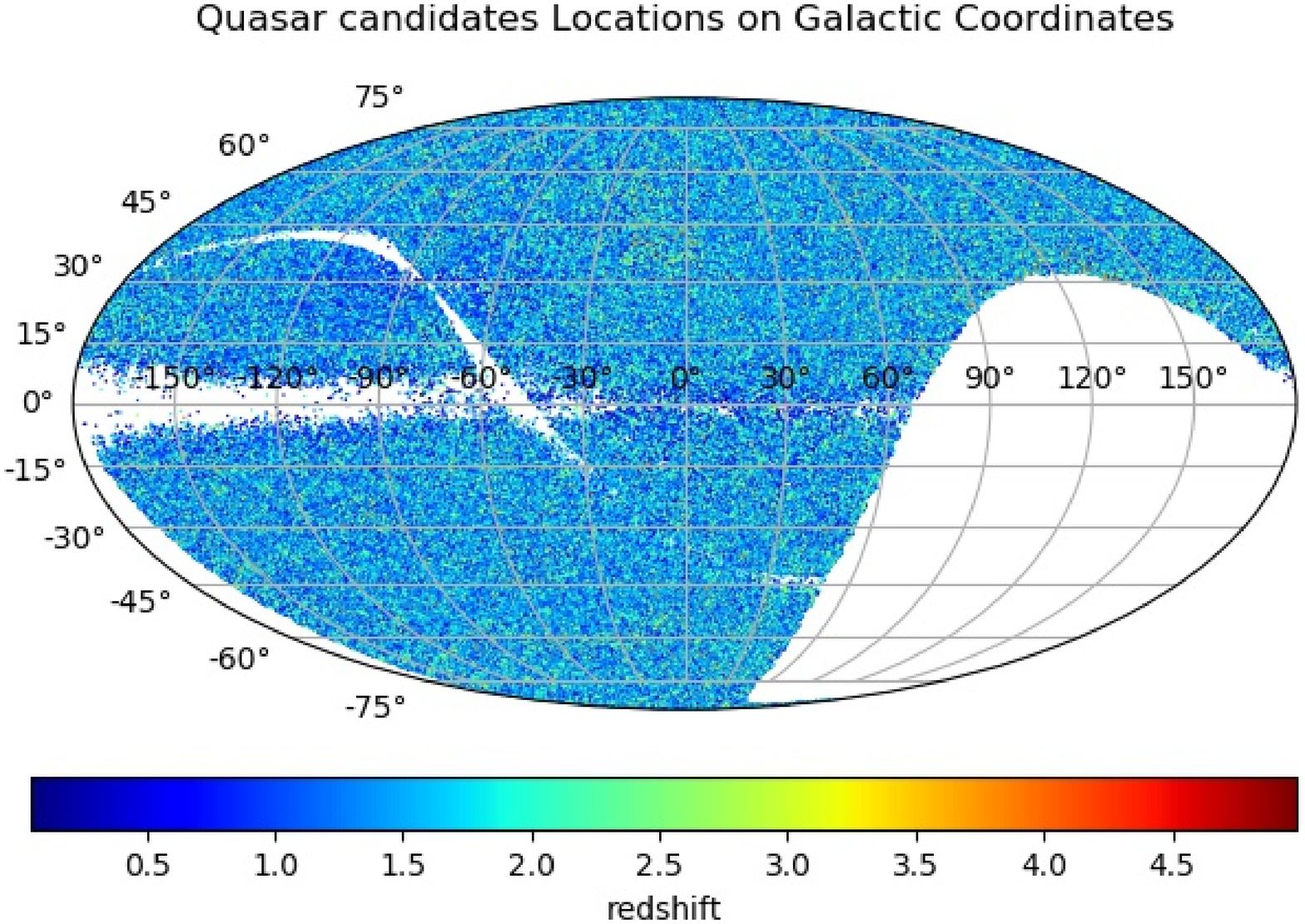}
\caption{The distribution of quasar candidates with $P_{\mathrm {QSO}}>0.95$ on Galactic locations}
\label{1}
\end{figure}

\subsection{Application of Quasar Candidates}

Through the above experiments, Table~7 further indicates those quasar candidates by XGB$\cap$SVM in different probabilities and redshift intervals.
Therefore we may choose appropriate quasar candidates for follow-up observation according to the observation condition and telescope sites.

The Large Sky Area Multi-Object Fibre Spectroscopic Telescope (LAMOST; \citealt{2012RAA....12.1197C}) is a 4-meter class reflecting Schmidt telescope with 20 square degree field of view (FOV) and 4,000 fibres, located at the Xinglong Station of National Astronomical Observatories of Chinese Academy of Sciences. It has the ability to obtain spectra of celestial objects with great efficiency. LAMOST has launched five spectroscopic surveys up to now, and the sixth spectroscopic survey is in progress. In the past five years, about 50,000 quasars were identified. For more efficient identification of quasars, careful preparation of the quasar input catalogue for LAMOST is important.
In Table~7, we focus on the quasar candidates with the probability $P_{\mathrm {QSO}}>0.95$. These quasar candidates will be added into the LAMOST quasar input catalogue and greatly enriches it.
Considering the magnitude limitation ($r$ magnitude less than 20) and telescope site (declination larger than $-10$), 269,121 candidates with $P_{\mathrm {QSO}}>0.95$ are kept.
Among them, 1,649 have predicted redshift ($z>3.5$), 40,624 have predicted redshift ($2.2<z\leq3.5$) and 226,848 quasars have redshift ($z\leq2.2$).

\section{Conclusion}
We cross-match the Pan-STARRS1 MeanObject table with AllWISE to obtain the sources containing optical and infrared photometric information. Then we put forward color criterions and a new schema to select quasar candidates. Our experimental results show that quasar candidates can be selected very efficiently by color criterions and machine learning algorithms based on optical and infrared photometric information. According to the color criterions (\texttt{yW1W2} and \texttt{iW1zW2}), most stellar pollution is excluded and most quasars are included. Therefore the optical and infrared color criterions are rather efficient and convenient methods to select quasar candidates. Although these techniques have such goodness, they don't consider all known information. If using all information or in higher dimensional spaces, the efficiency will improve. Therefore using all the features from optical and infrared bands as input patterns, we build two machine learning classifiers (XGBoost and SVM) to select quasar candidates. These two classifiers behave very well in the quasar and star classification, with accuracy of more than $99.46\%$ to the flux limits of the training set (see Figure 1). Applying the classifiers to the unknown cross-matched data set, a total of 2,006,632 sources ($P_{\mathrm{QSO}}>0.5$) are predicted to be quasar candidates. Among them, 1,657,988 have the probabilities $P_{\mathrm{QSO}}>0.75$ and 1,201,211 have much higher probabilities $P_{\mathrm{QSO}}>0.95$.

By comparing the performance of photometric redshift estimation of XGBoost with that of SVM, XGBoost is superior to SVM, at least to the flux limits of the training set. Then XGBoost is adopted as the core algorithm to predict photometric redshifts of the selected quasar candidates. Using the XGBoost regression, we estimate photometric redshift for each quasar candidate. In those quasars with high probability ($P_{\mathrm{QSO}}>0.95$), 7,402 are predicted to have high redshift ($z>3.5$), 230,181 quasars have medium redshift ($2.2<z\leq3.5$) and 963,628 quasars have low redshift ($z\leq2.2$).
Under the observation requirements of LAMOST ($r<20$, Dec $>-10$), 1,649 have predicted redshifts ($z>3.5$), 40,624 quasars with redshifts ($2.2<z\leq3.5$) and 226,848 quasars with redshifts ($z\leq2.2$).
These candidates will be added into the input catalogue of the LAMOST telescope to identify new quasars.

In summary, the color-cuts and the new scheme we put forward to pick out quasar candidates are reliable and efficient to the flux limits of the training set. The information added from infrared band is of great value to select quasar candidates, especially high redshift quasars. Our selected quasar candidates can be observed by the LAMOST telescope or other large telescopes. In the following research, we will consider time series information from GAIA, Pan-STARRS or future LSST to select quasar candidates.

\section*{Acknowledgements}
We are very grateful to the constructive comments and suggestions of Dr Adam David Myers, which help us improve our paper.
This paper is funded by 973 Program 2014CB845700 and the National Natural Science Foundation of China under grant Nos. 11873066 and U1731109.

We acknowledge the use of SDSS, PS1 and WISE photometric data.
Funding for the Sloan Digital Sky Survey IV has been provided by the Alfred P. Sloan Foundation, the U.S. Department of Energy Office of Science, and the Participating Institutions. SDSS-IV acknowledges
support and resources from the Center for High-Performance Computing at
the University of Utah. The SDSS web site is www.sdss.org.
SDSS-IV is managed by the Astrophysical Research Consortium for the
Participating Institutions of the SDSS Collaboration including the
Brazilian Participation Group, the Carnegie Institution for Science,
Carnegie Mellon University, the Chilean Participation Group, the French Participation Group, Harvard-Smithsonian Center for Astrophysics,
Instituto de Astrof\'isica de Canarias, The Johns Hopkins University,
Kavli Institute for the Physics and Mathematics of the Universe (IPMU) /
University of Tokyo, the Korean Participation Group, Lawrence Berkeley National Laboratory,
Leibniz Institut f\"ur Astrophysik Potsdam (AIP),
Max-Planck-Institut f\"ur Astronomie (MPIA Heidelberg),
Max-Planck-Institut f\"ur Astrophysik (MPA Garching),
Max-Planck-Institut f\"ur Extraterrestrische Physik (MPE),
National Astronomical Observatories of China, New Mexico State University,
New York University, University of Notre Dame,
Observat\'ario Nacional / MCTI, The Ohio State University,
Pennsylvania State University, Shanghai Astronomical Observatory,
United Kingdom Participation Group,
Universidad Nacional Aut\'onoma de M\'exico, University of Arizona,
University of Colorado Boulder, University of Oxford, University of Portsmouth,
University of Utah, University of Virginia, University of Washington, University of Wisconsin,
Vanderbilt University, and Yale University.

The Pan-STARRS1 Surveys (PS1) and the PS1 public science archive have been made possible through contributions by the Institute for Astronomy, the University of Hawaii, the Pan-STARRS Project Office, the Max-Planck Society and its participating institutes, the Max Planck Institute for Astronomy, Heidelberg and the Max Planck Institute for Extraterrestrial Physics, Garching, The Johns Hopkins University, Durham University, the University of Edinburgh, the Queen's University Belfast, the Harvard-Smithsonian Center for Astrophysics, the Las Cumbres Observatory Global Telescope Network Incorporated, the National Central University of Taiwan, the Space Telescope Science Institute, the National Aeronautics and Space Administration under Grant No. NNX08AR22G issued through the Planetary Science Division of the NASA Science Mission Directorate, the National Science Foundation Grant No. AST-1238877, the University of Maryland, Eotvos Lorand University (ELTE), the Los Alamos National Laboratory, and the Gordon and Betty Moore Foundation.

This publication makes use of data products from the Wide-field Infrared Survey Explorer, which is a joint project of the University of California, Los Angeles, and the Jet Propulsion Laboratory/California Institute of Technology, funded by the National Aeronautics and Space Administration.
%%%%%%%%%%%%%%%%%%%%%%%%%%%%%%%%%%%%%%%%%%%%%%%%%%

%%%%%%%%%%%%%%%%%%%% REFERENCES %%%%%%%%%%%%%%%%%%

% The best way to enter references is to use BibTeX:

\bibliographystyle{mnras}
\bibliography{bibfile} % if your bibtex file is called example.bib

% Alternatively you could enter them by hand, like this:
% This method is tedious and prone to error if you have lots of references
%\begin{thebibliography}{99}
%\bibitem[\protect\citeauthoryear{Author}{2012}]{Author2012}
%Author A.~N., 2013, Journal of Improbable Astronomy, 1, 1
%\bibitem[\protect\citeauthoryear{Others}{2013}]{Others2013}
%Others S., 2012, Journal of Interesting Stuff, 17, 198
%\end{thebibliography}

%%%%%%%%%%%%%%%%%%%%%%%%%%%%%%%%%%%%%%%%%%%%%%%%%%

%%%%%%%%%%%%%%%%% APPENDICES %%%%%%%%%%%%%%%%%%%%%

\appendix

% Don't change these lines
\bsp	% typesetting comment
\label{lastpage}
\end{document}